\documentclass[%
 reprint,
 amsmath,amssymb,
 aps,
]{revtex4-1}

\usepackage{graphicx}
\usepackage{dcolumn}
\usepackage{bm}

\usepackage{lipsum}
\usepackage{url}
\usepackage{hyperref}

\usepackage{multirow}

\makeatletter
\newcommand*{\rom}[1]{\expandafter\@slowromancap\romannumeral #1@}
\makeatother

\begin{document}

\preprint{APS/123-QED}

\title{Prospect of Chandrasekhar's limit against modified dispersion relation}

\author{Arun Mathew}
\email{a.mathew@iitg.ac.in}
\author{Malay K. Nandy}%
 \email{mknandy@iitg.ac.in}
\affiliation{Department of Physics, Indian Institute of Technology Guwahati, Guwahati 781039, India
}%

\date{August 9, 2019}

\begin{abstract}

Newtonian gravity predicts the existence of white dwarfs with masses far exceeding the Chandrasekhar limit when the equation of state of the degenerate electron gas incorporates the effect of quantum spacetime fluctuations (via a modified dispersion relation) even when the strength of the fluctuations is taken to be very small. In this paper, we show that this Newtonian ``super-stability'' does not hold true when the gravity is treated in the general relativistic framework. Employing dynamical instability analysis, we find that the Chandrasekhar limit can be reassured even for a range of high strengths of quantum spacetime fluctuations with the onset density for gravitational collapse practically remaining unaffected.  

\end{abstract}

\maketitle

\section{Introduction} 

The Hawking-Wheeler foam \cite{Wheeler1964,Hawking1978} of quantum space-time fluctuations can be accounted for by a non-commutative spacetime geometry. This modifies the dispersion relation between energy and momentum of any particle. Since a modification in the dispersion relation leads to a modified equation of state (EoS), it is apparent that the stellar structure of white dwarfs governed by its electron degenerate gas undergoes a measurable change if the stiffness of the EoS changes sufficiently.  In fact it has been shown \cite{Bertolami2010, Mathew2018b} that white dwarfs with modified EoS can support masses much higher than the Chandrasekhar limit that become ``super-stable'' when the hydrostatic equilibrium is governed by Newtonian gravity.

However the possibility of the existence of excessively massive white dwarfs is unlikely as it is inconsistent with an extensive amount of astronomical observations \cite{Shipman1972, Shipman1977, Shipman1979, Vennes1997, Marsh1997, Vennes1999,  Kilic2007}. Although there has been speculations about super Chandrasekhar white dwarf, It has argued \cite{Hicken2007,Silverman2011} that they are in fact double degenerate merges of two sub Chandrasekhar white dwarf. It is thus extremely important to investigate whether the above-mentioned ``super-stability'' prevails in general relativity (GR). It may be recalled that general relativity has a profound effect in determining the stability of a massive white dwarf against gravitational collapse although it has an insignificant effect on the stellar structure. 

We thus anticipate that a dynamical instability would set in at a critical value of the central density as generally predicted for relativistic stars \cite{Chandrasekhar1964a,Chandrasekhar1964b,Chandrasekhar1964c, Bardeen1966,Wheeler1968,Thorne1969}. Following this standard method employed earlier, we calculate the eigenfrequencies of the normal mode of radial oscillations with respect to various central densities of white dwarfs with the electron degenerate gas treated in the framework of modified dispersion relation. The existence of a vanishing eigenfrequency corresponds to the maximum central density in stable configuration. Consequently, we identify the onset density of dynamical instability with respect to a parameter characterizing the strength of quantum spacetime fluctuations. 

The notion of noncommutative geometry was introduced by Snyder \cite{Snyder1947a, Snyder1947b} by regarding the space-time coordinates as Hermitian operators preserving Lorentz invariance and incorporating a characteristic length scale. Heisenberg-like commutation relations between position and momentum do not hold in this framework and they resemble the generalized uncertainty relations explored from different points of view \cite{Mead1964, Veneziano1986, Padmanabhan1987, Amati1989, Konishi1990, Greensite1991, Maggiore1993, Kempf1995, Adler1999, Pedram2012a}. Moreover related viewpoints have emerged from various other considerations about the quantum nature of spacetime \cite{Rovelli1991a, Rovelli1991b, Amelinocamelia1994, Amelinocamelia1996}. In particular, Amelino-Camelia \cite{Amelinocamelia1997} argued that the $\kappa$-deformed Poincar\'e algebra, namely,  $[x^i,x^j]=0$ and $[x^i, t] =  ix^i/\kappa$ lead to similar conclusions. 

Amelino-Camelia \cite{Amelinocamelia2002}  further proposed a modified dispersion relation of the form $E^2-p^2c^2+f(E,p,m,\ell_P)=m^2c^4$ on the basis of a modified special relativity incorporating the Planck scale $\ell_{\rm P}$. Such dispersion relations have been utilized in different contexts. For example,  Alexander and Magueijo \cite{Alexander2004} addressed the horizon and flatness problems in a Friedmann cosmology with a modified dispersion relation of the form $E^2-p^2c^2f(E)=0$. Bertolami and Zarro \cite{Bertolami2010} considered a similar deformed dispersion relation, namely, $E^2=p^2c^2(1+\lambda E)^2+m^2c^4$ to address its effect on the the stability of astrophysical objects. Their calculation indicated that the stability of white dwarfs would be enchanced due to the increase in the pressure of the electron degenerate gas. 

In this paper, we use the modified dispersion relation $E_{\textbf p}^2=\mathbf{p}^2c^2(1+\lambda E_{\textbf p})^2+m^2c^4$ that leads to a modified equation of state of the degenerate electron gas. We analyze the stability of white dwarfs by calculating the eigenfrequencies of normal modes of small radial oscillations in the first order of perturbation. We find that general relativity is capable of causing a gravitational collapse even for high strengths of quantum spacetime fluctuations characterized by the parameter $\alpha=\lambda m_e c^2$. However when this strength is very high ($\alpha>0.0037$), the quantum space-time fluctuations become strong enough to hold up against any gravitational collapse. But the physical value of $\alpha$ is expected to be much lower than $0.0037$ for which we find that the Chandrasekhar critical limit can be realized in the quantum gravitational regime. 

The remainder of the paper is organized as follows. In Section \ref{Review}, we present a brief review of dynamical instability in general relativity. In Section \ref{NC_WD} the dynamical instability is explored for white dwarfs with the equation of state governed by a modified dispersion relation to account for the effect of quantum spacetime fluctuations on the instability. Finally we conclude the paper in Section \ref{Conclusion}.

\section{Dynamical instability in GR: A brief review}\label{Review}

In this section we review the dynamical instability of spherically symmetric fluid masses with respect to radial oscillations with the gravity treated general relativistically. The stability of the equilibrium fluid configuration is examined by considering radial perturbation in a spherically symmetric manner. Thus the metric interior to the fluid mass undergoing radial perturbation, given by
\begin{equation}\label{metric}
ds^2 = e^{\nu+\delta \nu} dt^2 - e^{\mu +\delta \mu}dr^2 -r^2(d\theta^2 + \sin^2 \theta \ d\phi^2),
\end{equation}
where $\nu(r)$ and $\mu(r)$ refer to the equilibrium configuration, and $\delta \nu(r,t)$ and $\delta \mu(r,t)$ correspond to a small radial Lagrangian displacement $\zeta(r,t)$ about the equilibrium configuration. Various quantities in the Einstein field equations, such as the pressure and energy density, become dependent on the coordinate time $t$ in addition to the the radial coordinate $r$. Supposing all perturbations are sinusoidal in the coordinate time, one can express the Lagrangian displacement as  $\zeta(r,t) = r^{-2} e^{\nu/2} \psi(r) e^{i\omega t}$ in the first order of perturbation. The equation governing the radial pulsation was obtained by Chandrasekhar \cite{Chandrasekhar1964a} which can be expressed in the Strum-Liouville form \cite{Bardeen1966}
\begin{equation}\label{main}
\frac{d}{dr}\left(U\frac{d\psi}{dr}\right) + \left(V + \frac{\omega^2}{c^2} W\right)\psi = 0,
\end{equation}
where 
\begin{eqnarray}
U(r) = \ &&e^{(\mu+3\nu)/2} \frac{\gamma P}{r^2}, \label{coefficient1} \\
V(r) = \ &&-4 \frac{e^{(\mu+3\nu)/2}}{r^3}\frac{dP}{dr} - \frac{8\pi G}{c^4} \frac{e^{3(\mu+\nu)/2}}{r^2}P(P+\varepsilon) \label{coefficient2} \nonumber \\
&&+ \frac{e^{(\mu+3\nu)/2}}{r^2} \frac{1}{P+\varepsilon} \left(\frac{dP}{dr}\right)^2,\\
W(r) = \ &&\frac{e^{(3\mu+\nu)/2}}{r^2}(P+\varepsilon), \label{coefficient3}
\end{eqnarray}
with $\gamma(r)$ the adiabatic index, $P(r)$ and $\varepsilon(r)$ are the pressure and energy density of the fluid mass at equilibrium given by the Tolman-Oppenheimer-Volkoff equation \cite{Tolman1939, Oppenheimer1939}
\begin{equation}\label{}
\frac{dP}{dr} = - \frac{G}{c^2 r} (\varepsilon + P) \frac{(m + 4\pi P r^3/c^2)}{(r-2GM/c^2)}
\end{equation}
with
\begin{equation}\label{}
\frac{dm}{dr} = \frac{4\pi}{c^2} \varepsilon r^2. 
\end{equation}

The admissible radial motion requires the fluid element at the center to have a vanishing displacement $\zeta$ (and $d\zeta/dr$ be finite) leading to the condition
\begin{equation}\label{BC1}
\psi = 0 \hspace{0.5cm} {\rm at} \hspace{0.5cm}  r=0.
\end{equation}
Furthermore, the solution is also required to satisfy
\begin{equation}\label{BC2} 
\delta P =-e^{\nu/2}\frac{\gamma P}{r^2}\frac{d\psi}{dr}= 0 \hspace{0.5cm} {\rm at}\hspace{0.5cm}  r=R,  
\end{equation}
at the surface, where $\delta P$ represents the Lagrangian change in pressure and $R$ is the radius of the fluid sphere. 

Multiplying Eq.~(\ref{main}) from left by $\psi$ and integrating from $0$ to $R$ and applying the above boundary conditions, one can obtain the integral 
\begin{equation}\label{Rayleighritz}
J[\psi] = \int_0^R \left\{ U \psi'^2 -V\psi^2 -  \frac{\omega^2}{c^2} W \psi^2 \right\} dr. 
\end{equation}
where $\psi'=d\psi/dr$. Minimizing $J[\psi]$ with respect to $\psi$ yields the Strum-Liouville equation (\ref{main}), thus providing a variational basis for determining the lowest characteristic eigenfrequency given by 
\begin{equation}\label{Rayleighritz}
\frac{\omega_0^2}{c^2} =  \min_{\psi(r)} \frac{ \int_0^R \left\{ U \psi'^2 -V\psi^2 \right\} dr}{\int_0^R W \psi^2  dr},
\end{equation}
corresponding to the normal mode of oscillation. A sufficient condition for the dynamical instability to set in is that the value of $\omega_0^2$ be negative. Thus a suitable trial function $\psi(r)$ satisfying the boundary conditions and making the right hand side vanish will determine the onset of instability.  It can be seen that a power series solution for the above Strum-Liouville equation (\ref{main}) about $r=0$ satisfying the boundary conditions (\ref{BC1}) and (\ref{BC2}) has a leading order term $\propto r^3$. To determine the eigenfrequency of the fundamental mode we assume a trial function $\psi(r)\propto r^3$ which corresponds to a homologous vibration. It can be shown \cite{Mestel1966} that this trial function is a sufficiently close approximation to the true eigenfunction of the fundamental mode. 

\section{Dynamical instability in white dwarfs with modified dispersion relation} \label{NC_WD}

It has been reported that when quantum spacetime fluctuations are included in the equation of state of the electron gas, white dwarfs can exist in excessively large masses beyond the Chandrasekhar limit when the gravity is treated in the Newtonian framework although the strength of quantum spacetime fluctuations is taken to be very small \cite{Rashidi2016, Mathew2018a}. However, it is well-known that, in the conventional problem of stability of white dwarfs, a dynamical instability sets in \cite{Chandrasekhar1964b, Chandrasekhar1964c} when the gravity is treated in the framework of general relativity. It is thus natural to speculate that a similar general relativistic instability may set in when the small effect of quantum spacetime fluctuations is included in the equation of state of the electron gas. 

\subsection{Modified equation of state} 

The effect of quantum spacetime fluctuations can be  modelled via a modified dispersion relation. We shall take modified dispersion relation \cite{Bertolami2010}
\begin{equation}
E_{\textbf p}^2=\mathbf{p}^2c^2(1+\lambda E_{\textbf{p} })^2+m^2c^4
\end{equation}
which imposes a momentum cutoff at $p_{\rm max}=(\lambda c)^{-1}$ above which the energy becomes unphysical. For small values of momentum it coincides with the ideal dispersion relation $E_{\textbf p}^2=\mathbf{p}^2c^2+m^2c^4$, but it deviates strongly for high values of momentum, becoming infinite at the cutoff $p_{\rm max}$. 

Since the electron gas in white dwarfs is completely degenerate, we evaluate the pressure $P$, internal energy $\varepsilon_{\rm int}$,  and mass-density $\rho_0 $ at absolute zero from the grand partition fucntion. Thus we obtain \cite{Mathew2018b} the modified equation of state
\begin{equation}\label{EOS1}
P = A \tilde{P}(\xi) , \hspace{0.2cm} \rho_0 c^2 = \frac{A}{q} \xi^3, \hspace{0.2cm} \varepsilon = \rho_0 c^2 + \varepsilon_{\rm int} = \frac{A}{q} \tilde{\varepsilon}(\xi)
\end{equation}
where 
\begin{equation}\label{DEOS}
\tilde{P}(\xi) = \xi^3 f(\xi)-3g(\xi), \hspace{0.5cm} \tilde{\varepsilon} = (1-q)\xi^3 + \ 3q g(\xi), 
\end{equation}
\begin{equation}\label{EOS2}
\xi=\frac{p_{F}}{m_e c}, \hspace{0.2cm} A = \frac{8\pi m_e^4 c^5}{3 h^3},   \hspace{0.2cm}   q = \frac{m_e}{\mu_e m_u} = 2.74297\times10^{-4},
\end{equation}
\begin{equation}\label{f}
f(\xi) =\left[ \alpha \xi^2+ \sqrt{(1-\alpha^2)\xi^2+1} \right] \biggl[1-\alpha^2\xi^2 \biggr]^{-1},
\end{equation}
and
\begin{multline}\small\label{g}
g(\xi)= \frac{1}{\alpha^4}\biggl[ 2\tanh^{-1}{\alpha \xi}+\tanh^{-1}{\frac{\xi(1-\alpha^2)}{\alpha+\sqrt{1+(1-\alpha^2)\xi^2}}} \\
 -\frac{(2-\alpha^2)}{2\sqrt{1-\alpha^2}}\sinh^{-1}{\xi\sqrt{1-\alpha^2}} \biggr] \\
 - \frac{\xi}{3\alpha^3}\left(3+\alpha^2\xi^2+\frac{3\alpha}{2}\sqrt{1+(1-\alpha^2)\xi^2}\right) 
\end{multline}
with $\alpha = \lambda m_e c^2$. 

Employing the relativistic expression $\gamma = \frac{\tilde{\varepsilon}+q\tilde{P}}{\tilde{P}} \left( \frac{d\tilde{P}}{d\tilde{\varepsilon}} \right)_s$ for the adiabatic index $\gamma$, and using Eqs.~(\ref{EOS1}\---\ref{g}), we obtain
\begin{equation} 
\gamma = \frac{\tilde{\varepsilon}(\xi)+q\tilde{P}(\xi)}{\tilde{P}(\xi)} \left(\frac{\xi}{3}\right)\left(1-q+q\frac{dg}{d\xi}\right)^{-1} \left(\frac{df}{d\xi}\right) 
\end{equation}
for the electron gas with the modified equation of state.

\subsection{Stability Analysis} 

The Einstein field equation for the static interior Schwarzschild metric can be solved \cite{Tolman1939, Oppenheimer1939} to obtain the metric components
\begin{equation}\label{lambda}
e^{-\mu} = 1-2q\frac{\tilde{m}}{\eta}
\end{equation}
and 
\begin{equation}\label{nu1}
e^\nu = \left(1-2q \frac{\tilde{M}}{\eta_R}\right) \times \exp\left\{ -2q\int_0^\eta \frac{\tilde{P}'}{(\tilde{\varepsilon}+q\tilde{P})}d\eta\right\},
\end{equation} 
where $\tilde{m}=m/m_0$ and $\eta=r/r_0$ are dimensionless variables with 
\begin{equation}
r_0 = \frac{qc^2}{\sqrt{4\pi A G}}  \hspace{0.5cm} {\rm and} \hspace{0.5cm} m_0 = \frac{q^2c^4}{\sqrt{4\pi A}} \frac{1}{G^{3/2}}.
\end{equation}
The dimensionless quantities $\tilde{M}=M/m_0$ and $\eta_R=R/r_0$ correspond to the mass $M$ and radius $R$ of the fluid sphere. Using Eqs.~(\ref{DEOS}\---\ref{g}), Eq.~(\ref{nu1}) can be simplified to
\begin{equation}\label{nu2}
e^\nu =\left(1-2q \frac{\tilde{M}}{\eta_R}\right) \left(\frac{1}{1-q+q f(\xi)}\right)^2. 
\end{equation}

The dependence of the above field quantities on the radial coordinate $\eta$ correspond to the Tolman-Oppenheimer-Volkoff equation \cite{Tolman1939,Oppenheimer1939} of hydrostatic equilibrium expressed here as 
\begin{equation}\label{dPdr}
 \frac{d\tilde{P}}{d\eta}   = \left(\frac{df}{d\xi}\right)\xi^3 \frac{d\xi}{d\eta}= -\left(\frac{\tilde{\varepsilon}+q\tilde{P}}{\eta}\right)\left(\frac{\tilde{m}+ q\tilde{P}\eta^3 }{\eta-2q\tilde{m}}\right)
\end{equation}
with the mass equation 
\begin{equation}
\frac{d\tilde{m}}{d\eta} = \tilde{\varepsilon} \eta^2.
\end{equation}

The functions $U(\eta)$, $V(\eta)$ and $W(\eta)$ in the Strum-Liouville equation (\ref{main}) are readily obtained from Eqs.~(\ref{coefficient1}\---\ref{coefficient3}) employing Eqs.~(\ref{lambda}), (\ref{nu2}) and (\ref{dPdr}). In consequence, Eq.~(\ref{Rayleighritz}) yields the eigenfrequency 
\begin{equation}\label{omega}
\omega_0^2 = \left(\frac{q c^2}{r_0^2}\right)  \frac{ \mathcal{I} +  \mathcal{J}}{\mathcal{K}},
\end{equation}
where 
\begin{equation}\label{I}
\mathcal{I}=\int_0^{\eta_R} e^{(\mu+3\nu)/2} \frac{\gamma \tilde{P}}{\eta^2} \psi_0'^2  d\eta,
\end{equation}
\begin{eqnarray}\label{J}
\mathcal{J}=\int_0^{\eta_R}\frac{e^{(\mu+3\nu)/2}}{\eta^2}\left[\frac{4}{\eta} \frac{d\tilde{P}}{d\eta} + 2q e^\mu \tilde{P}(\tilde{\varepsilon}+q\tilde{P}) \right. \nonumber\\
\left. -\frac{q}{\tilde{\varepsilon}+q\tilde{P}} \left(\frac{d\tilde{P}}{d\eta}\right)^2\right] \psi_0^2 d\eta,  
\end{eqnarray}
and
\begin{equation}\label{K}
\mathcal{K}=\int_0^{\eta_R} e^{(3\mu+\nu)/2} \frac{\tilde{\varepsilon}+q\tilde{P}}{\eta^2} \psi_0^2 d\eta,
\end{equation}
with $\psi_0$ the eigenfunction associated with the fundamental mode that minimizes the right-hand side of Eq.~(\ref{Rayleighritz}). 

\begin{figure}
\centering
\includegraphics[width=8cm]{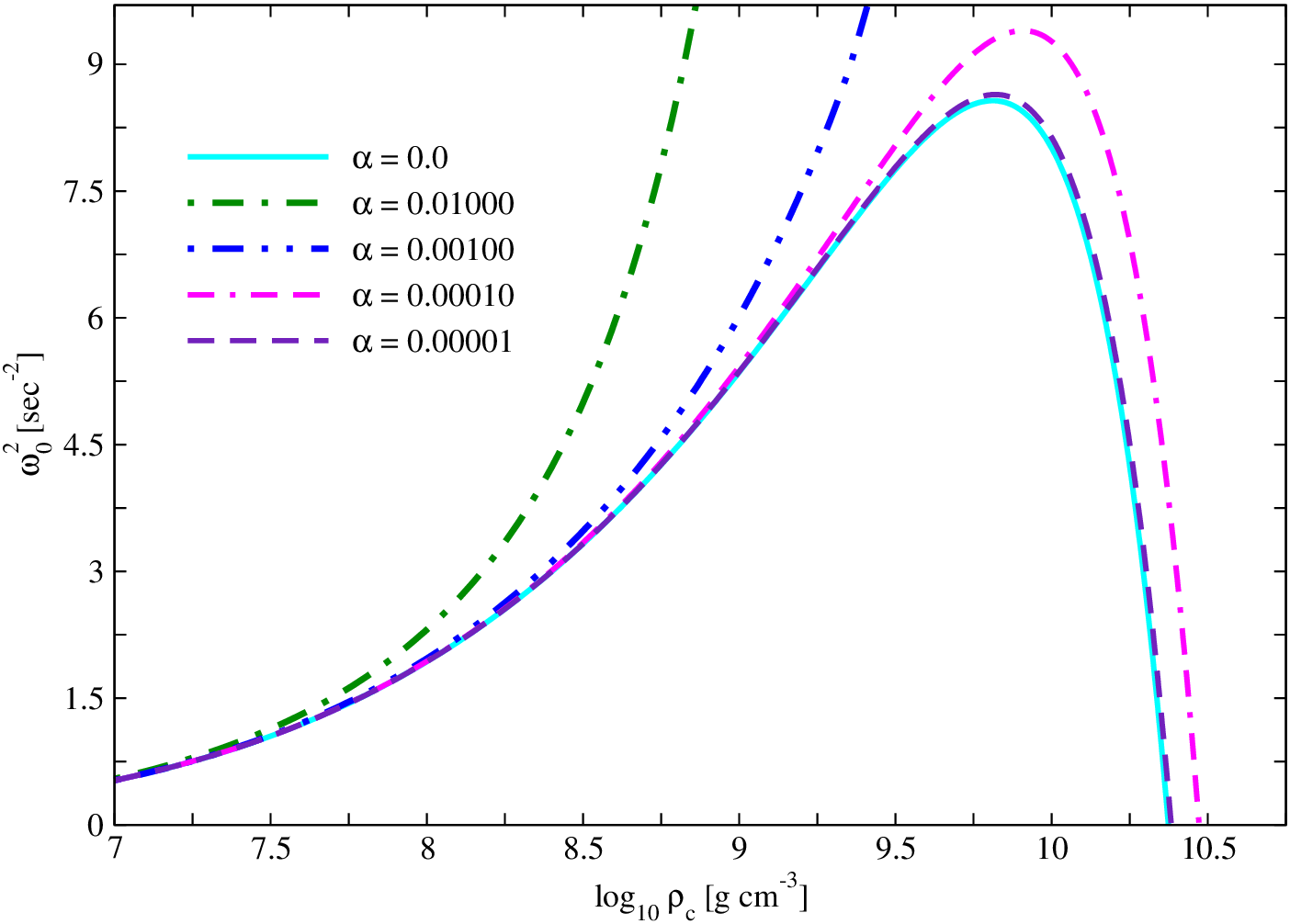}
\caption{Eigenfrequency for normal modes against central density of relativistic white dwarfs for various values of the parameter $\alpha$ characterising the strength of quantum spacetime fluctuations.}
\label{Figure_1}
\end{figure}

As stated earlier, we make the choice $\psi_0=\eta^3$ as a trial function and  evaluate the integrals $\mathcal{I}$, $\mathcal{J}$, and $\mathcal{K}$ given by Eqs.~(\ref{I}\---\ref{K}) for different values of the central Fermi momentum $\xi_c$. The corresponding eigenfrequencies are obtained from Eq.~(\ref{omega}). Figure \ref{Figure_1} displays the eigenfrequency against the central density $\rho_c$ [related to $\xi_c$ through $\rho = (A/qc^2) \{  (1-q)\xi^3 + \ 3q g(\xi) \}$] for different strengths of quantum spacetime fluctuations parametrized by $\alpha$, namely, $\alpha = 0.01$, $0.001$, $0.0001$, and $0.00001$, including the ideal case ($\alpha=0$).  

We observe from Figure \ref{Figure_1} that for $\alpha=0.00001$, the characteristic eigenfrequencies are close to the ideal values. As the strength is increased to $\alpha=0.0001$, the eigenfrequencies depart from the ideal values but they follow a trend similar to the ideal case, and gravitational instability can set in dictated by general relativity by virtue of the existence of a vanishing eigenfrequency and a corresponding critical central density $\rho_c^*$. This signifies the dominance of  gravitational pull determined by general relativity over the effect of quantum spacetime fluctuations on the equation of state. 

\begin{figure}
\centering
\includegraphics[width=8cm]{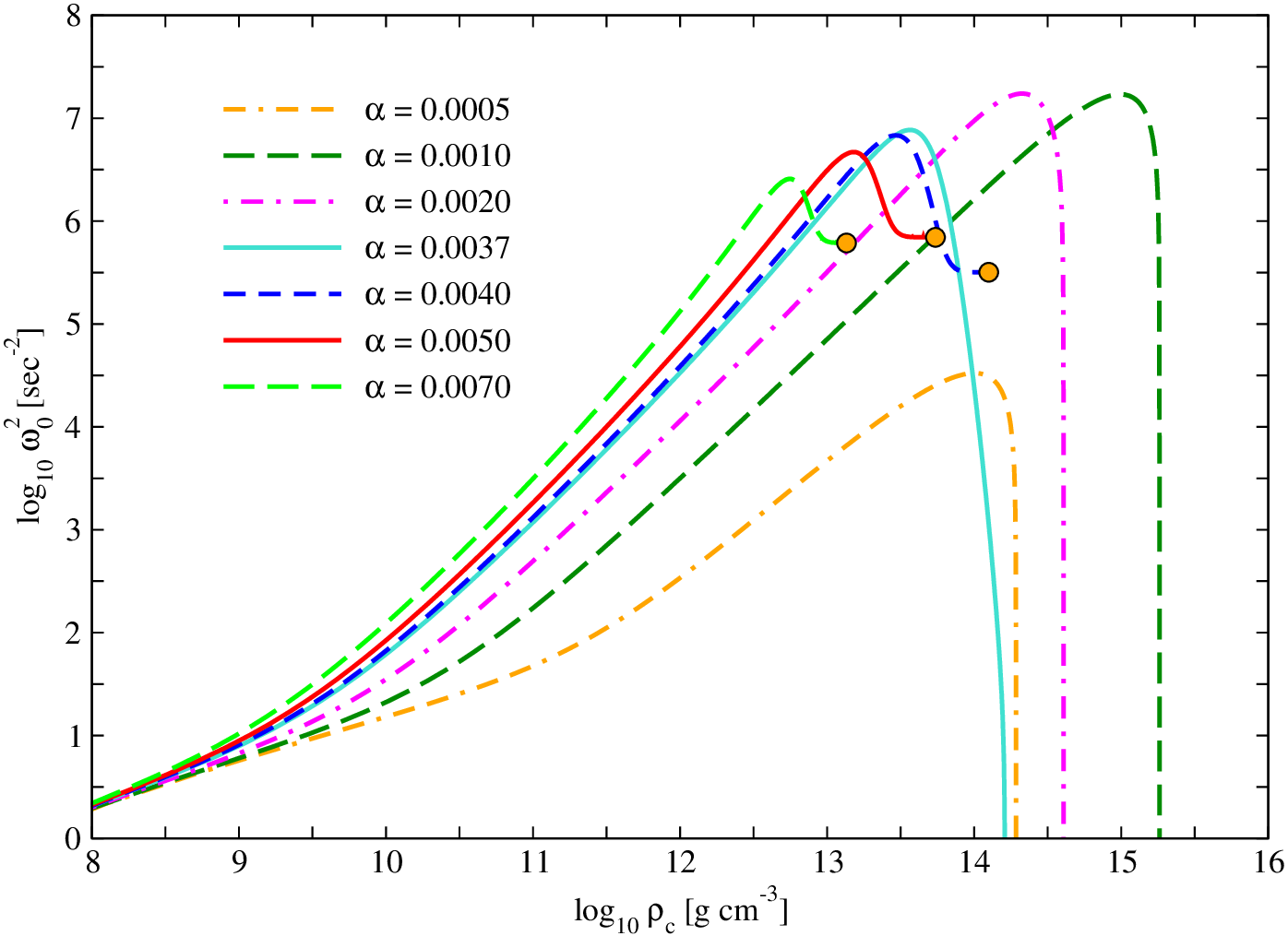}
\caption{Eigenfrequency for normal modes versus central density of relativistic white dwarfs for various strengths of $\alpha$.}
\label{Figure_2}
\end{figure}

However, for higher strengths of $\alpha$, such as $\alpha=0.001$ and $0.01$, the scale of the ordinate in Figure \ref{Figure_1} is not adequate to analyze their behaviors. For an adequate analysis of the situation, we show a log-log plot in Figure \ref{Figure_2}. It is clear from Figure \ref{Figure_2} that for $\alpha\leq0.0037$, the trends of the eigenfrequencies are similar to that of the ideal case and general relativistic instabilities can set in at critical values of the central densities $\rho_c^*$ due to the existence of vanishing eigenfrequencies of the normal mode. 

On the other hand, for $\alpha>0.0037$, the curves follow a trend completely different from the ideal case because they reach terminal points at non-zero eigenfrequencies and thus zero eigenfrequency solutions do not exist. This signifies the non-existence of any gravitational instability or collapse and the white dwarfs remain ``super-stable''. The central densities at which the curves terminate are higher than  Chandrasekhar's value of $2.3\times 10^{10}$ g cm$^{-3}$. From the super-stability for $\alpha>0.0037$, we may say that quantum spacetime fluctuations become sufficiently strong so that the gravitational pull determined by general relativity is incapable of bringing about any instability. 
 
\begin{figure}
\centering
\includegraphics[width=8cm]{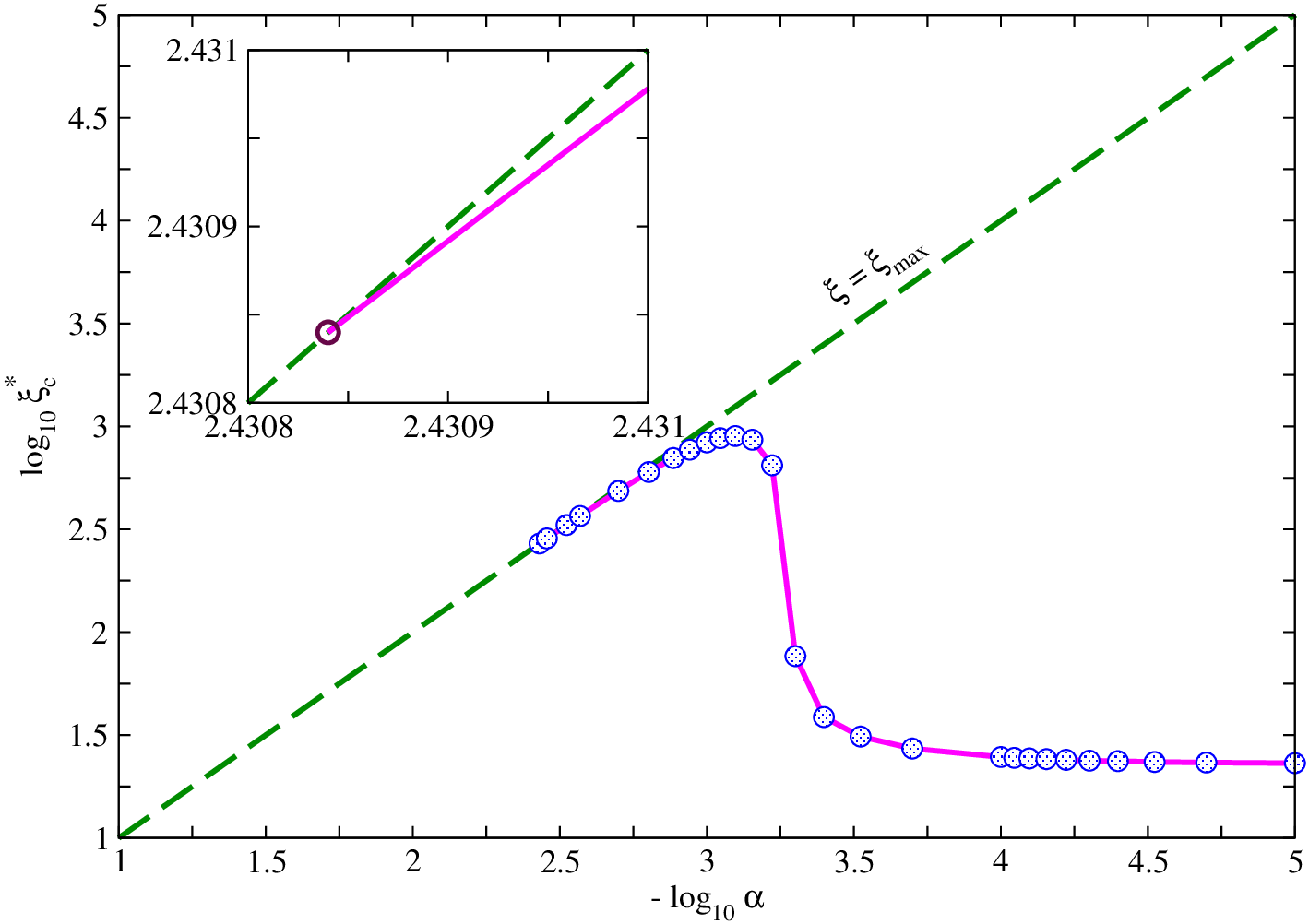}
\caption{Critical value of the central Fermi momentum for the onset of gravitational instability versus the parameter $\alpha$. The dashed straight line represents $\xi=\xi_{\rm max} = \alpha^{-1}$. The inset shows the intersection of the two curves at $\alpha=0.0037$.}
\label{Figure_3}
\end{figure}

Since the general relativistic instability (corresponding to the zero eigenfrequency) occurs at a critical central density $\rho_c^*$ determined by the strength of $\alpha$, it is worth studying the behavior of the critical values of the central Fermi momenta with respect to the parameter $\alpha$. We see from the right-hand part of Figure \ref{Figure_3} that as the strength of $\alpha$ is increased from $10^{-5}$ to $10^{-4}$, the critical value $\xi_c^*$ (or equivalently $\rho_c^*$) remains approximately constant. In fact, there is negligible variation in the value of $\xi_c^*$ in the range $0<\alpha<10^{-5}$ (which is also evident from Figure \ref{Figure_1} from the near-coincidence of the two $\rho_c^*$ values). Thus in the range $0<\alpha<10^{-4}$, we expect that general relativistic instability would yield approximately equal critical masses. In this range, we find $\rho_c^* = 2.3\--2.9\times 10^{10}$ g cm$^{-3}$ which is in the vicinity of Chandrasekhar's value of $2.3\times 10^{10}$ g cm$^{-3}$ \cite{Chandrasekhar1964c} . This indicates that Chandrasekhar's general relativistic critical mass of $1.42$ M$_\odot$ is hardly affected in this range of the parameter $\alpha$. 

We thus see that the effect of general relativity is robust enough to cause an instability against the effect of quantum spacetime fluctuations even for strengths such as $\alpha=10^{-4}$ which is in fact a very large value. If the strength of $\alpha$ is increased beyond $10^{-4}$, we see from Figure \ref{Figure_3} that the critical central Fermi momentum $\xi_c^*$ also increases but eventually falls off and approaches the curve $\xi = \alpha^{-1}= \xi_{\rm max}$ (the straight line) until it makes an intersection at $\alpha=0.0037$, the maximum strength of $\alpha$ for the existence of a vanishing eigenfrequency. This intersection is shown in the inset of Figure \ref{Figure_3} by an open circle where $\xi_c^*= (0.0037)^{-1}$. It is obvious that the curve cannot cross the line $\xi=\alpha^{-1}$, having reached the maximum value $\xi_c^* =\alpha^{-1} $. 

\section{Conclusion}\label{Conclusion}

Since the effect of quantum gravity is inevitably present  \cite{Wheeler1964, Hawking1978}, such effects, however small, must be taken into account when the density of matter is sufficiently high. In this context it is worthwhile to recall that the standard Chandrasekhar limit of $1.44$ M$_\odot$ is approached when the density of the electron degenerate gas in white dwarfs approaches infinity in the framework of Newtonian gravity. The problem thus calls for taking account of the effect of quantum space-time fluctuations at ultra-high densities of the electron degenerate gas. When this notion is followed and the electron gas is treated via a modified dispersion relation making the equation of state more stiff than the ideal one, it is found that white dwarfs become ``super-stable'' and higher masses beyond the Chandrasekhar limit are possible \cite{Mathew2018b} when the gravity is treated in the Newtonian framework. However, as noted in the introduction, white dwarfs are found to exist only below the Chandrasekar limit. It is thus extremely important to resolve this paradoxical situation.

It is known from Chandrasekhar's study \cite{Chandrasekhar1964a, Chandrasekhar1964c} that a dynamical instability sets in when the gravity is treated general relativistically in white dwarfs. Consequently it is natural to ask the question whether general relativity would be capable of reassuring the Chandrasekar limit when the effect of quantum space-time fluctuations is included in the equation of state.

Motivated by the above query, we analyzed the problem of stability of white dwarfs governed by general relativity and incorporating quantum space-time fluctuations in the electron degenerate gas via a modified equation of state.

To analyze the stability, we followed the standard methodology of perturbations generating radial pulsations in spherically symmetric white dwarfs and calculated the corresponding eigenfrequencies of the radial oscillations. The corresponding eigenvalue equation is in the Strum-Liouville form whose eigenfrequencies can be related to a variational principle. With an appropriate trial function for the Lagrangian displacement, we calculated the eigenfrequencies for various strengths of the quantum space-time fluctuations parametrized by $\alpha$. 

We find that, for large values of $\alpha$ such that $\alpha>0.0037$, white dwarfs remain ``super-stable'' as they do not exhibit any zero eigenfrequency in the normal mode. Such white dwarfs can support maximum masses determined by the maximum values of the central density $\rho_c$ where the curves terminate. These values of $\rho_c$ are higher than the critical density obtained by Chandrasekhar suggesting the possibility of white dwarfs of masses higher than Chandrasekar's general relativistic value of $1.42$ M$_\odot$. However these cases are unphysical because we do not expect that the strength $\alpha$ of spacetime fluctuations to be as large as $0.0037$ or higher. 

On the other hand, for smaller strengths, such that $\alpha\leq0.0037$, we find that general relativity is capable of bringing about an instability at finite central densities $\rho_c^*$ because of the existence of a vanishing eigenfrequency in the normal mode. This signifies that gravity governed by general relativity is strong enough to cause a gravitational collapse against the effect of quantum spacetime fluctuations on the equation of state. It is important to note that general relativity is robust enough to bring about a gravitational collapse even for very high strengths of quantum spacetime fluctuations such as $10^{-3}$ or $10^{-4}$. In fact these values of $\alpha$ are expected to be unphysically high for quantum spacetime fluctuations. 

We have also seen from Figures \ref{Figure_1} and \ref{Figure_3} that in the range $0<\alpha<10^{-4}$,  general relativistic instability yields comparable critical central densities, namely, $\rho_c^* = 2.3\--2.9\times 10^{10}$ g cm$^{-3}$. This range is in the vicinity of Chandrasekhar's value of $2.3\times 10^{10}$ g cm$^{-3}$ \cite{Chandrasekhar1964c}. This indicates that the stellar structure of relativistic white dwarfs is hardly affected in this range of the parameter $\alpha$ where the critical mass is about $1.42$ M$_\odot$.

We may recall that when the gravity is treated in the Newtonian framework, masses far exceeding the Chandrasekhar limit are found to be super-stable even for very low values of $\alpha$. It is thus obvious that while Newtonian gravity is unable to dominate over the stiffness of the equation of state generated by quantum spacetime fluctuations, general relativity does possess the capacity to overcome the stiffness of the equation of state that can lead to a gravitational collapse. 

Thus even for a high value of $\alpha$, such as $10^{-4}$ or $10^{-5}$, the onset density $\rho_c^{*}$ for gravitational collapse is practically unaffected (with respect to the ideal case) when the gravity is treated general relativistically in spite of the effect of quantum spacetime fluctuations opposing gravitational collapse. This is of direct relevance  to the core collapse supernovae where the degenerate core of the progenitors are found to have a mass of about $1.4$ M$_\odot$. However it may be recalled that there may be no clear distinction between the gravitational core collapse and fast $\beta$-capture that may occur nearly simultaneously at the onset, effectively making no difference in the impending supernova explosion \cite{Wheeler1968}. Thus our study suggests that when the inevitable effect of quantum spacetime fluctuations is included in the process, the situation practically remains indistinguishable from the ideal core collapse scenario .  

\section*{Acknowledgement} 
Arun Mathew is indebted to the Ministry of Human Resource Development, Government of India, for financial assistance through a doctoral fellowship.


%

\end{document}